\documentclass{Interspeech}

\usepackage{cite}
\usepackage{subcaption}
\usepackage{times}
\usepackage{epsfig}
\usepackage{graphicx}
\usepackage{amsmath}
\usepackage{amssymb}
\usepackage{listings}
\usepackage{xcolor}
\usepackage{fancybox}
\usepackage{hyperref}
\usepackage[many]{tcolorbox}

\interspeechcameraready

\title{Room Impulse Response as a Prompt for Acoustic Echo Cancellation}

\author[affiliation={1}]{Fei}{Zhao}
\author[affiliation={1}]{Shulin}{He}
\author[affiliation={1}]{Xueliang}{Zhang}

\affiliation{College of Computer Science}{Inner Mongolia University}{China}

\email{\{zhaofei, heshulin\}@mail.imu.edu.cn, cszxl@imu.edu.cn}
\keywords{acoustic echo cancellation, room impulse response prompt, deep learning}

\usepackage{comment}

\begin{document}

\maketitle

\begin{abstract}
    
    Data-driven acoustic echo cancellation (AEC) methods, predominantly trained on synthetic or constrained real-world datasets, encounter performance declines in unseen echo scenarios, especially in real environments where echo paths are not directly observable. Our proposed method counters this limitation by integrating room impulse response (RIR) as a pivotal training prompt, aiming to improve the generalization of AEC models in such unforeseen conditions.  
We also explore four RIR prompt fusion methods. Comprehensive evaluations, including both simulated RIR under unknown conditions and recorded RIR in real, demonstrate that the proposed approach significantly improves performance compared to baseline models. These results substantiate the effectiveness of our RIR-guided approach in strengthening the model's generalization capabilities. The official code and the audio samples are available at https://github.com/ZhaoF-i/RIR-Prompt-AEC.
\end{abstract}

\section{Introduction}
\label{sec:intro}
Acoustic Echo Cancellation (AEC) is a critical research area in speech signal processing, playing a vital role in hands-free communication systems \cite{sondhi1967adaptive,benesty2001advances,enzner2014acoustic}. The primary goal of AEC is to eliminate echo from the microphone signal using a far-end reference signal \cite{4648922, 10096597, Zhao2024}. Traditional approaches predominantly rely on adaptive filtering \cite{paleologu2015overview, haykin2005adaptive, bershad1979analysis, kuech2014state} to model the echo path from the far-end signal to the microphone. These methods offer advantages such as minimal near-end speech distortion and robust performance across diverse acoustic conditions. However, their effectiveness is significantly compromised in scenarios involving nonlinear echoes or double-talk, resulting in residual echoes that degrade communication quality \cite{guerin2004nonlinear, halimeh2019neural, patel2024nonlinear, yin2024nonlinear}. This issue is further aggravated by the use of low-quality amplifiers and speakers.

As deep learning technologies advance, data-driven approaches to AEC have gradually become mainstream. Zhang et al. \cite{DBLP:conf/icassp/ZhangLZ22, DBLP:journals/taslp/ZhangLL023} proposed in-place convolution recurrent neural networks (ICRN), which utilize in-place convolution and channel-wise temporal modeling for preserving the near-end signal information. Zhang et al. \cite{zhang2022multi} proposed MTFAA, multi-scale time-frequency processing and streaming axial attention-based approach. However, these methods heavily depend on the availability of large datasets to ensure the robust performance of the models. The challenge of model generalization becomes especially pronounced when the models are deployed in real-world environments where the data may significantly deviate from that observed during training \cite{9413585, 9746272, Panchapagesan2022, sridhar2021icassp, saka2023conversational}. This discrepancy underscores an enduring challenge within the field, as the models must contend with the variability inherent in real-world data.

More recently, hybrid approaches to AEC have risen to prominence, skillfully combining the strengths of traditional algorithms with the power of deep learning to overcome individual shortcomings \cite{revach2021kalmannet,yang2023low,DBLP:journals/corr/abs-2301-12363}. They demonstrate superior echo suppression and robust performance in complex acoustic environments, surpassing the capabilities of conventional techniques. Additionally, hybrid methods provide improved speech quality preservation and enhanced model generalization compared to deep learning approaches.
Although the compact design of hybrid AEC models optimizes computational efficiency, it can also restrict the network's capacity, potentially capping performance enhancements. 
This trade-off becomes particularly apparent in low Signal-to-Echo Ratio (SER) scenarios.


This paper presents an innovative approach to augment the generalization capabilities of AEC network models by employing Room Impulse Response (RIR) as a prompt. The procedure begins with a communication device emitting a pulse signal in a noisy environment, which is then captured by the device's microphone. These signals, integrated with the far-end and the microphone signals, serve as the network's input. It is crucial to note that the initial pulse signal collection is performed preemptively, with subsequent recollection triggered by hardware-detected changes in the echo path, such as those indicated by device gyroscope movements, thereby ensuring the method's practical applicability. 
Further, we have investigated four innovative fusion techniques leveraging RIR as input data. We have selected the ICCRN \cite{liu2023iccrn} and the MTFAA \cite{zhang2022multi} as our AEC baseline models of choice. Through extensive testing that encompasses mismatched synthetic data and real-world RIR scenarios, we demonstrate that our method markedly enhances model generalization. These findings substantiate the effectiveness of employing RIR as a prompt in AEC models.

\begin{figure}[t!]
        \centering
	\includegraphics[width=1\linewidth]{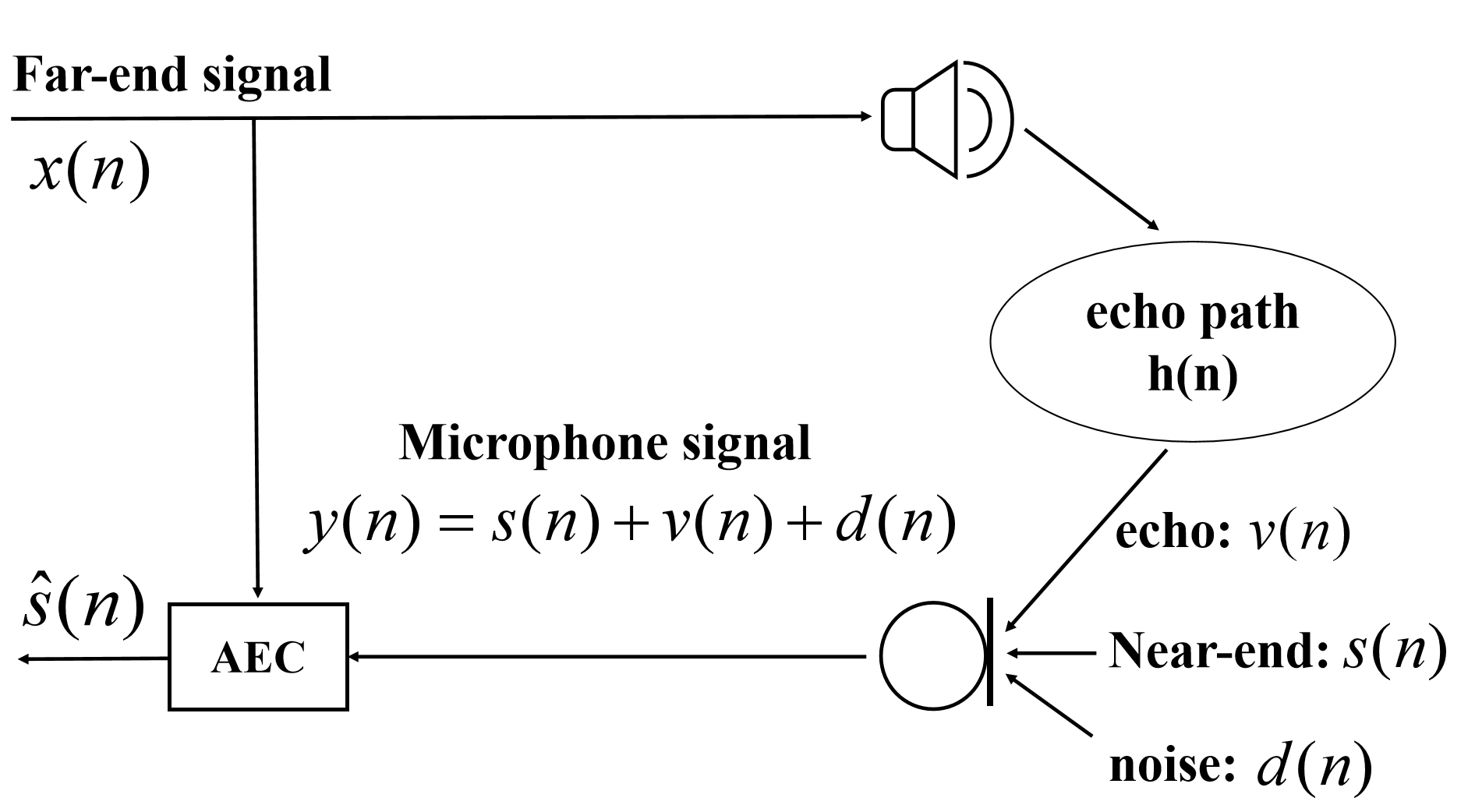}
	\caption{Diagram of the single channel AEC.}
        \vspace{-0.5cm}
	\label{fig:AEC}
\end{figure}


\begin{figure*}
        \centering
	\includegraphics[width=\textwidth]{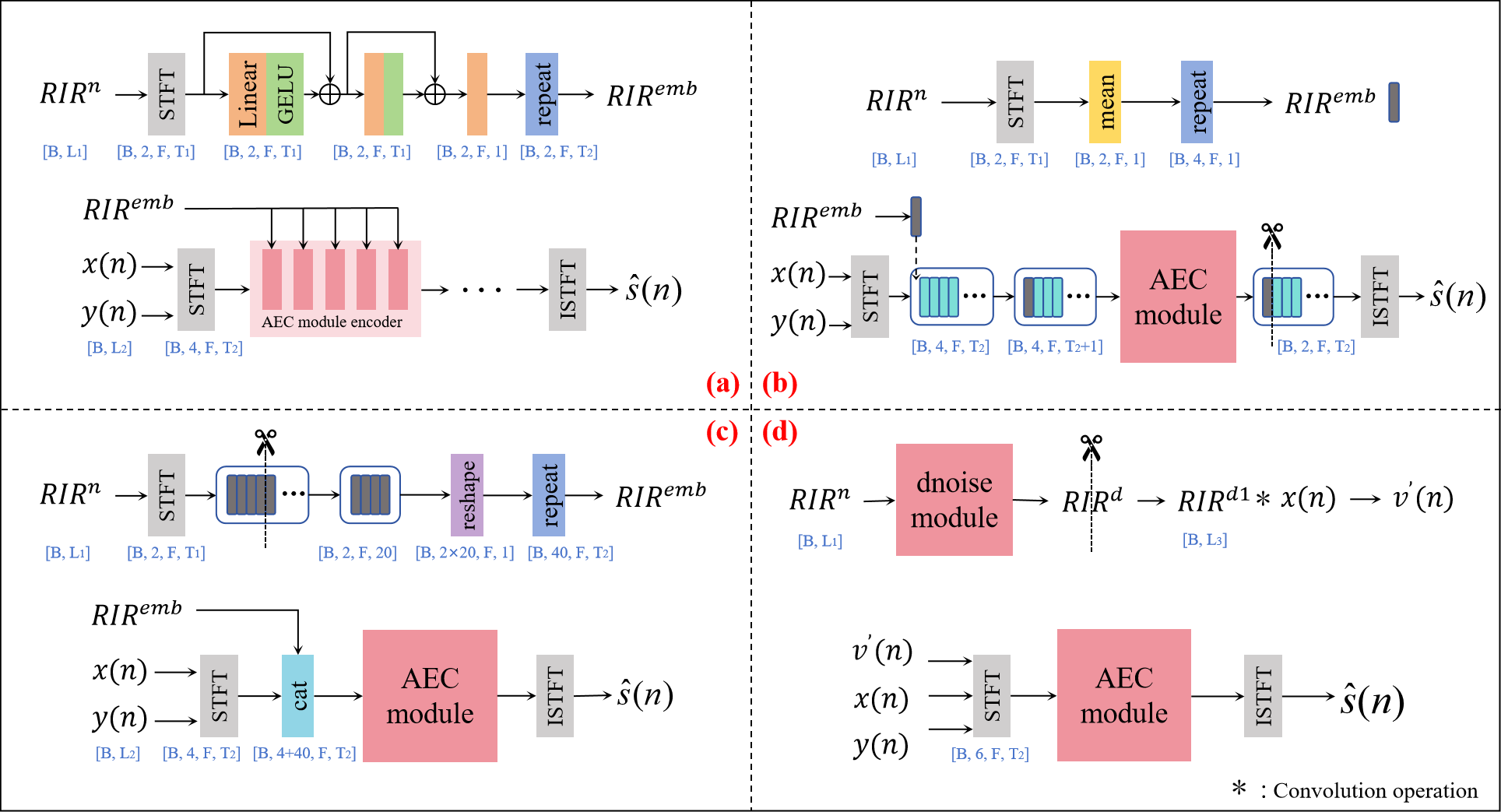}
	\caption{Four different fusion methods for RIR prompt.}
        \vspace{-0.2cm}
	\label{fig:fusion}
\end{figure*}

\begin{figure}
        \centering
	\includegraphics[width=1\linewidth]{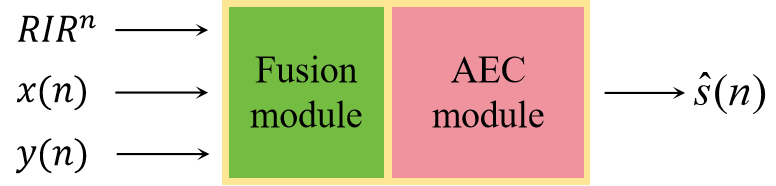}
	\caption{Overview of proposed method.}
        \vspace{-0.5cm}
	\label{fig:overview}
\end{figure}
The rest of this paper is organized as follows. In Section 2,
the problem is formulated. In Section 3, we describe the proposed
method. The experimental setup is shown in Section 4. In Section 5, we demonstrate the performance of the proposed system. We conclude the paper in Section 6.

\section{Problem Formulation}
\label{sec:FORMULATION}

The single-channel AEC is shown in \autoref{fig:AEC}, the system typically has access to two critical input signals: the near-end microphone signal $y(n)$ and the far-end signal $x(n)$, $n$ represents the time sampling point. The formulation of the microphone signal in the time domain is presented as follows:
\begin{equation}
     y(n) = s(n) + v(n) + d(n)
\end{equation}
where the acoustic echo $v(n)$ is $x(n)$ convolving with a room impulse response (RIR) \cite{habets2006room} $h(n)$, $s(n)$ is the near-end speech signal and $d(n)$ is a possible near-end noise signal.


\section{Proposed Method}
\label{sec:METHOD}

\subsection{Overview}
\label{ssec:Overview}
In this section, we provide an overview of our proposed method, as illustrated in \autoref{fig:overview}. The network incorporates the noisy RIR signal (denoted as ${RIR}^n$) as a complementary input to the AEC model. The RIR data is first preprocessed by a dedicated fusion module and then merged with the regular input of the AEC model for further processing.

\subsection{Fusion module}
\label{ssec:Fusion}

\autoref{fig:fusion} demonstrates an exploration of four unique fusion techniques that employ RIR as a prompt signal. In Figure 2(a), the noisy RIR signal ${RIR}^n$ undergoes a Short-Time Fourier Transform (STFT) to transition from the time domain to the time-frequency (T-F) domain. It then sequentially traverses two linear layers followed by a Gaussian Error Linear Unit (GELU) activation function, employing residual connections to mitigate information loss. These linear layers independently model the frequency and time dimensions, after which the signal is subjected to another linear layer that compresses the time dimension from $T_1$ to a single unit, facilitating temporal aggregation. Post the repeat operation, a prompt embedding ${RIR}^{emb}$ is derived, which scales the time dimension from 1 to $T_2$, aligning with the temporal requirements for the subsequent concatenation operation with the AEC model's input signal. For the final merging step, we concatenate ${RIR}^{emb}$ with the input of each layer encoder along the channel dimension.

In Figure 2(b), our approach commences with the application of STFT to the ${RIR}^n$. This is followed by a temporal averaging step and channel repetition, which serve to condense the signal information into a single, comprehensive frame. This frame, encapsulating the essential features, is then concatenated with both the far-end and near-end microphone signals within the T-F domain. Following its passage through the AEC module, the initial frame is discarded.

In Figure 2(c), ${RIR}^n$ is processed through STFT and then cropped to 20 frames, thereby reducing the number of subsequent channels. Subsequently, the signal is reshaped from format $[B, 2, F, 20]$ to $[B, 2\times20, F, 1]$, followed by a repeat operation that extends the time dimension to $T_2$, resulting in ${RIR}^{emb}$. Finally, ${RIR}^{emb}$ is concatenated with the far-end and microphone signals in the T-F domain along the channel dimension. This signal is then forwarded to the AEC module for further processing.

Figure 2(d) illustrates the final fusion technique, a streamlined approach to information utilization. Initially, ${RIR}^n$ is processed by a rudimentary denoising module, yielding the denoised signal ${RIR}^d$. This module employs a masking technique for denoising, a strategy chosen for its ability to minimize computational complexity. Importantly, it is designed to retain as much RIR information as possible, thereby enhancing the fidelity of the signal. 
To mitigate the computational complexity in subsequent steps, ${RIR}^d$ is strategically truncated to create ${RIR}^{d1}$. Subsequently, ${RIR}^{d1}$ is convolved with the far-end signal $x(n)$ to derive the prompt echo signal $v'(n)$. The echo signal, along with the far-end and microphone signals, is used as input for the AEC module to predict the near-end signal. 



\subsection{Loss function}
\label{sssec:Loss-function}
In this method, the corresponding loss function consists of multiple items. Stretched Scale-Invariant Signal-to-Noise Ratio (S-SISNR)\cite{DBLP:conf/interspeech/SunYZH21} is a modified version of the Scale-Invariant Signal-to-Noise Ratio (SISNR)\cite{DBLP:journals/taslp/LuoM19} loss function. S-SISNR is a time domain loss function that is obtained by doubling the period of SISNR. The simplified formula for S-SISNR is expressed as follows:
\begin{equation}
	\mathcal{L}_{\text {s-sisnr }}=10 \log_{10} cot^2(\frac{\beta}{2}) = 10\log_{10}\frac{1+cos(\beta)}{1-cos(\beta)}
\end{equation}
where $\beta$ represents the angle between two vectors' ideal near-end signal $s$ and predicted near-end signal $\hat{s}$, since it is complicated to calculate the half angle, after the derivation of the trigonometric function, it can be represented by $cos(\beta)$.

The “RI+Mag” loss criterion is adopted to recover the complex spectrum, following \cite{10447755, zhao2024attention}. In detail, the loss is defined as:
\begin{equation}
	\mathcal{L}_{\mathrm{mag}}=\frac{1}{TF} \sum_t^T \sum_f^F ||S(t, f)|^p-|\hat{S}(t, f)|^p|^2 
\end{equation}

\begin{equation}
	\mathcal{L}_{\mathrm{RI}}=\frac{1}{TF} \sum_t^T \sum_f^F ||S(t, f)|^p e^{j \theta_{S(t, f)}}-|\hat{S}(t, f)|^p e^{j \theta_{\hat{S}(t, f)}}|^2
\end{equation}
where $p$ is a spectral compression factor (set to 0.5) \cite{li2021importance}.
Operator $\theta$ calculates the phase of a complex number. Then the total loss function is as follows:
\begin{equation}
	\mathcal{L}_{total}=\mathcal{L}_{\mathrm{RI}}+\mathcal{L}_{\text{mag }}+\mathcal{L}_{\text{s-sisnr}}
\end{equation}

\section{Experimental Setups}
\label{sec:EXPERIMENTAL}

\subsection{Datasets}
The near-end and far-end signals employed in our experiments are sourced from the ICASSP 2023 AEC challenge's synthetic datasets, representing near-end and far-end scenarios, respectively \cite{AEC-challange}. The RIR is generated using the Image method \cite{allen1979image}. We simulate different rooms of size $l \times w \times h $ $m^3$ for training mixtures, where $l$ ranges from 4 to 8m, $w$ from 4 to 7, and $h$ from 3 to 5, each incremented by 1m. To emphasize the effectiveness of RIR prompts, the microphone-loudspeaker (M-L) distance is fixed at 0.3m. The reverberation time (T60) is randomly selected from [0.1, 0.2, 0.3, 0.4, 0.5, 0.6]s to generate RIRs in each room. Then echo speech $v(n)$ is mixed with near-end speech $s(n)$ at signal-to-echo ratio (SER) randomly chosen from [-10, 10]dB with step 1dB. Noises are selected from a sound effect library (available at https://www.sound-ideas.com) and added to the speech at signal-to-noise ratios in [5, 15]dB. The nonlinearity setting adheres to the configuration established in \cite{DBLP:conf/icassp/ZhangLZ22}. We have generated a dataset comprising 200,000 training samples, 10,000 validation samples, and 1,000 test samples via a randomized selection process, with non-linear settings accounting for $90\%$. All speech samples have a duration of 5 seconds and the sampling rate is 16kHz. 

To evaluate the generalizability of the proposed method, we established two supplementary test sets of 1000 each. The first set involved selecting the M-L distance for RIR generation from [0.4, 0.5, 0.8]m. The second set incorporated real RIRs obtained from actual acoustic environments \cite{DBLP:conf/icdsp/JeubSV09}.

\subsection{Training details}
For $L_1$ and $L_3$ in \autoref{fig:fusion}, we set them as 8000 and 3200 sampling points, respectively.
For the STFT used to calculate the “RI+Mag” loss ($\mathcal{L}_{\mathrm{mag}}$, $\mathcal{L}_{\mathrm{RI}}$), the window size was set to 20 ms with an offset of 5 ms using a hamming window. For the baseline models, both the ICCRN and the MTFAA, we have adopted a uniform configuration: a window length of 20 ms and a window shift of 10 ms for processing the signals. Additionally, we maintain consistency in the training loss function, training strategy, and the dataset utilized for training both models.
Furthermore, given that our training data operates at a sampling rate of 16 kHz, we have made the decision to exclude Modules band merging and band splitting from the MTFAA model. 
These models are optimized by Adam algorithm \cite{DBLP:journals/corr/KingmaB14}.
The initial learning rate is set to 0.001. If the validation loss does not decrease for two consecutive epochs, the learning rate is reduced by half. Training is stopped when the verification loss does not decrease for 10 consecutive epochs.

\begin{table*}[htbp]
  \centering
  \caption{Comparison of ICCRN and fusion of different RIR prompts ‘DT’ and ‘ST\_FE’ represent double talk and far-end single talk scenarios respectively. For each metric, bold indicates the best value.}
    \scalebox{1.1}{
    \fontsize{9}{10}\selectfont
    \begin{tabular}{lccccccccc}
    \toprule
     Test RIR types     & \multicolumn{3}{c}{Match Synthetic RIR}  & \multicolumn{3}{c}{Mismatch Synthetic RIR}   & \multicolumn{3}{c}{Real RIR} \\
    \cmidrule(lr){2-4}\cmidrule(lr){5-7}\cmidrule(lr){8-10}
    
    Test scenarios & \multicolumn{2}{c}{DT}  & \multicolumn{1}{c}{ST\_FE} & \multicolumn{2}{c}{DT} & \multicolumn{1}{c}{ST\_FE} & \multicolumn{2}{c}{DT}  & \multicolumn{1}{c}{ST\_FE} \\

    \cmidrule(lr){2-3}\cmidrule(lr){4-4}\cmidrule(lr){5-6}\cmidrule(lr){7-7}\cmidrule(lr){8-9}\cmidrule(lr){10-10}
    
    Model\textbackslash Metric     & \multicolumn{1}{l}{PESQ} & \multicolumn{1}{l}{SDR} & \multicolumn{1}{l}{ERLE} & \multicolumn{1}{l}{PESQ} & \multicolumn{1}{l}{SDR} & \multicolumn{1}{l}{ERLE} & \multicolumn{1}{l}{PESQ} & \multicolumn{1}{l}{SDR} & \multicolumn{1}{l}{ERLE} \\
    \midrule
    mix     & 1.71    & -1      & \multicolumn{1}{c}{—} & 1.71    & -1      & \multicolumn{1}{c}{—} & 1.76    & -1      & \multicolumn{1}{c}{—} \\
    \midrule
    ICCRN   & 2.96    & 13.34   & 43.08   & 1.98    & 4.28    & 5.32    & 2.05    & 3.61    & 3.73 \\
    + RIR prompt fusion (a) & 3.01    & 14.66   & \textbf{47.13}   & 1.94    & 3.7     & 4.6     & 2.08    & 4.02    & 3.54 \\
    + RIR prompt fusion (b) & 2.97    & 14.01   & 42.44   & 1.95    & 4.03    & 5.23    & 2.13    & 4.29    & 3.99 \\
    + RIR prompt fusion (c) & 2.99    & \textbf{14.78}   & 45.95   & 1.96    & 3.84    & 4.29    & 2.11    & 4.57    & 4.00 \\
    + RIR prompt fusion (d) & \textbf{3.03}    & 13.23   & 44.94   & \textbf{2.12}    & \textbf{6.13}    & \textbf{10.7}   & \textbf{2.19}    & \textbf{4.64}    & \textbf{4.79} \\
    \midrule
    MTFAA   & 2.46    & 11.56   & 25.43   & 2.43    & 11.32    & 24.51    & 2.03    & 7.75    & 9.79 \\
    + RIR prompt fusion (d) & \textbf{2.59}    & \textbf{12.14}   & \textbf{29.37}   & \textbf{2.56}    & \textbf{11.98}     & \textbf{28.48}     &\textbf{ 2.25}    & \textbf{10.16}    & \textbf{16.1} \\
    \bottomrule
    \end{tabular}%
    }
  \label{tab:compare}%
   \vspace{-0.1cm}
\end{table*}%

\begin{table}
\centering
\caption{The computational amount and parameter amount of the two baseline models and the RIR prompt fusions.}
\scalebox{1.1}{
\begin{tabular}{lcc}
\toprule
Model & Macs(G) & Param(M)    \\
\midrule
ICCRN  & 1.93    & 0.463       \\
+ RIR prompt fusion (a)      & 1.94    & 0.495       \\
+ RIR prompt fusion (b)      & 1.95    & 0.464       \\
+ RIR prompt fusion (c)      & 2.05    & 0.499       \\
+ RIR prompt fusion (d)      & 2.77   & 0.611       \\
\midrule
MTFAA  & 5.41   & 2.149       \\
+ RIR prompt fusion (d)      & 6.26   & 2.269       \\
\bottomrule
\end{tabular}
}
\label{tab:mac}%
\vspace{-0.25cm}
\end{table}

\section{Experimental Results}
\label{sec:RESULTS}

To evaluate the performance of the proposed method, we utilize the echo return loss enhancement (ERLE) \cite{rix2001perceptual}, the perceptual evaluation of speech quality (PESQ) \cite{enzner2014acoustic} and signal-to-distortion ratio (SDR) \cite{vincent2006performance} as the metrics that measure the echo suppression for single-talk, near-end speech quality and near-end speech fidelity for double-talk periods, respectively. Higher scores indicate better performance.

As shown in \autoref{tab:compare}, we validated our results across three distinct test sets to assess the robustness of the proposed method under various conditions. The first set, 'Match Synthetic RIR' mirrors the training set conditions, providing a baseline for comparison. The second set, 'Mismatch Synthetic RIR' introduces variations in the M-L distances to evaluate the model's adaptability to different acoustic configurations. The third set, 'Real RIR' incorporates actual recordings from real-world scenarios to test the method's performance in authentic environments. We focused on two specific scenarios: 'DT' for dual-talk situations, where both near-end and far-end participants speak simultaneously, and 'ST\_FE' for single-talk scenarios at the far-end, where only the far-end speaker is active. The RIR prompt fusion techniques (a), (b), (c), and (d) correspond to the distinct fusion methods illustrated in \autoref{fig:fusion}.

When evaluating the 'Match Synthetic RIR' scenario for the ICCRN model, it is observed that across all fusion methods leveraging RIR as a prompt, there is an enhancement in performance. Specifically, the fusion methods labeled (a), (c), and (d) each secured the highest score for one metric, thereby demonstrating the efficacy of the RIR prompt approach under test conditions that align with the training parameters.

In the 'Mismatch Synthetic RIR' test scenario for the ICCRN model, we observed that fusion methods (a), (b), and (c) did not yield performance improvements; in fact, they exhibited a decline in certain indicators. We postulate that this may be attributed to the utilization of noisy RIRs and potential limitations in the generalization capability of the ICCRN model. Conversely, fusion method (d) demonstrated a comprehensive advantage in this test set. This superior performance is likely due to the simple denoising process, which preserves more RIR information. Furthermore, the prompt echo derived from the convolution of the denoised RIR with the far-end signal allows for more direct utilization of the information.

In the final 'Real RIR' test scenario for the ICCRN model, we observed an enhancement in the performance of all four fusion methods. Notably, fusion method (d) demonstrated the most substantial improvement. This result suggests that method (d) is more adept at leveraging prompt information effectively. Furthermore, it substantiates the effectiveness of the proposed approach in enhancing the model's generalization capabilities.

Among the four fusion methods explored, method (d) in conjunction with the ICCRN demonstrated the most promising results. Consequently, we integrated the MTFAA model into the RIR fusion method (d) to further enhance performance. The results presented in the table indicate that the incorporation of RIR prompt information yields particularly notable improvements on the mismatched test set. Notably, the MTFAA model, when fused with prompt information, outperforms the ICCRN under the same conditions. We attribute this superior performance to the inherently larger computational scope and parameter count of the MTFAA model, which potentially allows for a more nuanced capture of the acoustic characteristics present in the mismatched test set.

In \autoref{tab:mac}, we present a comparative analysis of the computational complexity and parameter count between the two baseline models and their enhanced versions incorporating RIR prompt fusions. The results delineated in the table reveal that our proposed method introduces a modest increase in computational and storage demands, yet it yields a significant enhancement in performance.

\begin{figure}[t]
    \centering
    \begin{subfigure}[b]{0.15\textwidth}
        \includegraphics[width=\textwidth]{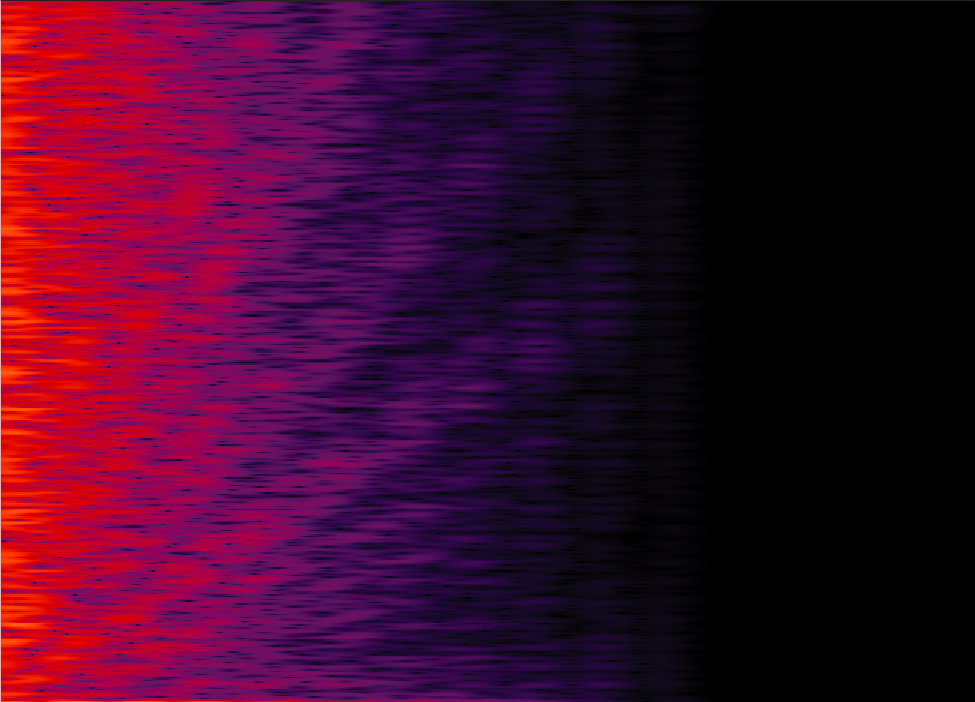}
        \caption{Clean RIR}
        \label{fig:sub1}
    \end{subfigure}
    \hfill
    \begin{subfigure}[b]{0.15\textwidth}
        \includegraphics[width=\textwidth]{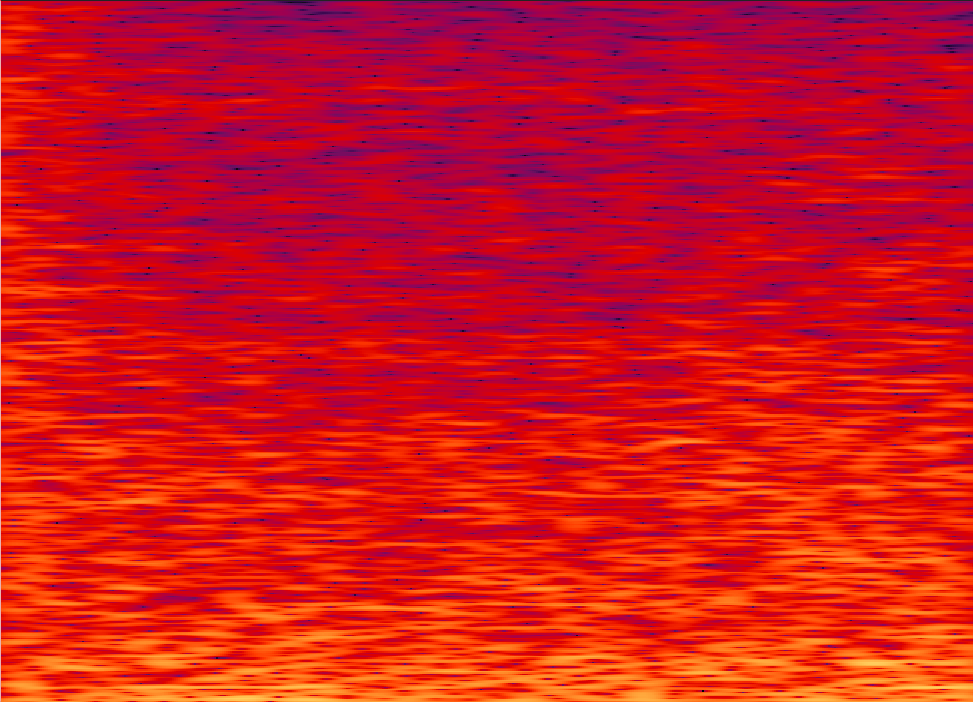}
        \caption{Noisy RIR}
        \label{fig:sub2}
    \end{subfigure}
    \hfill
    \begin{subfigure}[b]{0.15\textwidth}
        \includegraphics[width=\textwidth]{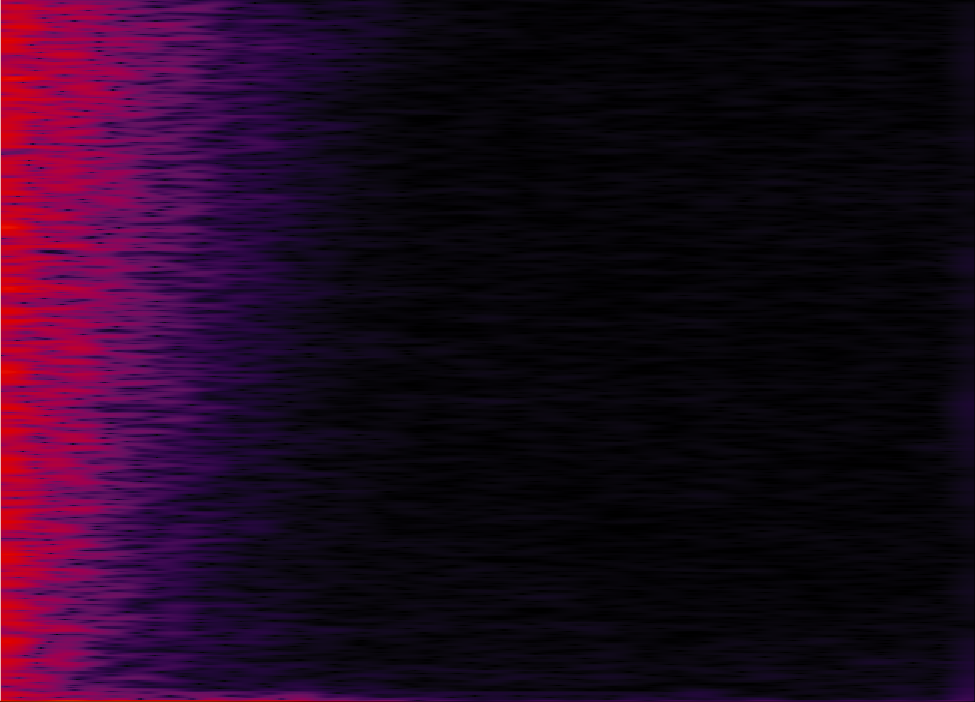}
        \caption{Denoise RIR}
        \label{fig:sub3}
    \end{subfigure}
    \caption{Spectrum of clean RIR, noisy RIR, and denoised RIR.}
    \label{fig:rir}
    \vspace{-0.35cm}
\end{figure}

Additionally, there are two reasons why traditional methods are not used: firstly, we assume that pulse signals are collected in a noisy environment, and for noisy RIR, as shown in \autoref{fig:sub2}, this signal is completely unusable for traditional methods. Secondly, collecting pulse signals in a noisy environment makes speech denoising an additional task in this study. Coupled with non-linear settings, the above limitations make traditional methods, even using denoising RIR (compare \autoref{fig:sub3} and \autoref{fig:sub1}) to initialize the filter, unable to handle this scenario well.

\section{Conclusions}
\label{sec:CONCLUSIONS}

In this paper, we propose a novel approach that utilizes room impulse response (RIR) as a prompt to enhance the generalization of the acoustic echo cancellation (AEC) model. We investigate four distinct methods for integrating RIR prompts. The effectiveness of our approach is validated across three test sets: synthetic RIR with matching conditions, synthetic RIR with mismatching conditions, and real-world RIRs. The results demonstrate that our method not only improves model performance under matching conditions but also enhances generalization in mismatching scenarios.

\textbf{Acknowledgments}: This research was partly supported by the China National Nature Science Foundation (No. 61876214) and CCF-Lenovo Research Fund (Grant No.20240203).

\bibliographystyle{IEEEtran}
\bibliography{mybib}

\begin{thebibliography}{10}
\providecommand{\url}[1]{#1}
\csname url@samestyle\endcsname
\providecommand{\newblock}{\relax}
\providecommand{\bibinfo}[2]{#2}
\providecommand{\BIBentrySTDinterwordspacing}{\spaceskip=0pt\relax}
\providecommand{\BIBentryALTinterwordstretchfactor}{4}
\providecommand{\BIBentryALTinterwordspacing}{\spaceskip=\fontdimen2\font plus
\BIBentryALTinterwordstretchfactor\fontdimen3\font minus \fontdimen4\font\relax}
\providecommand{\BIBforeignlanguage}[2]{{%
\expandafter\ifx\csname l@#1\endcsname\relax
\typeout{** WARNING: IEEEtran.bst: No hyphenation pattern has been}%
\typeout{** loaded for the language `#1'. Using the pattern for}%
\typeout{** the default language instead.}%
\else
\language=\csname l@#1\endcsname
\fi
#2}}
\providecommand{\BIBdecl}{\relax}
\BIBdecl

\bibitem{sondhi1967adaptive}
M.~Sondhi, ``An adaptive echo canceller,'' \emph{Bell System technical journal}, vol.~46, no.~3, pp. 497--511, 1967.

\bibitem{benesty2001advances}
J.~Benesty, T.~G{\"a}nsler, D.~R. Morgan, M.~M. Sondhi, S.~L. Gay \emph{et~al.}, ``Advances in network and acoustic echo cancellation,'' 2001.

\bibitem{enzner2014acoustic}
G.~Enzner, H.~Buchner, A.~Favrot, and F.~Kuech, ``Acoustic echo control,'' in \emph{Academic press library in signal processing}.\hskip 1em plus 0.5em minus 0.4em\relax Elsevier, 2014, vol.~4, pp. 807--877.

\bibitem{4648922}
C.~Paleologu, J.~Benesty, and S.~Ciochina, ``A variable step-size affine projection algorithm designed for acoustic echo cancellation,'' \emph{IEEE Transactions on Audio, Speech, and Language Processing}, vol.~16, no.~8, pp. 1466--1478, 2008.

\bibitem{10096597}
D.~Yang, F.~Jiang, W.~Wu, X.~Fang, and M.~Cao, ``Low-complexity acoustic echo cancellation with neural kalman filtering,'' in \emph{ICASSP}, 2023, pp. 1--5.

\bibitem{Zhao2024}
F.~Zhao, C.~Zhang, S.~He, J.~Liu, and X.~Zhang, ``Deep echo path modeling for acoustic echo cancellation.''\hskip 1em plus 0.5em minus 0.4em\relax International Speech Communication Association, 2024.

\bibitem{paleologu2015overview}
C.~Paleologu, S.~Ciochin{\u{a}}, J.~Benesty, and S.~L. Grant, ``An overview on optimized nlms algorithms for acoustic echo cancellation,'' \emph{EURASIP Journal on Advances in Signal Processing}, vol. 2015, pp. 1--19, 2015.

\bibitem{haykin2005adaptive}
S.~S. Haykin, \emph{Adaptive filter theory}.\hskip 1em plus 0.5em minus 0.4em\relax Pearson Education India, 2005.

\bibitem{bershad1979analysis}
N.~Bershad and P.~Feintuch, ``Analysis of the frequency domain adaptive filter,'' vol.~67, no.~12, pp. 1658--1659, 1979.

\bibitem{kuech2014state}
F.~Kuech, E.~Mabande, and G.~Enzner, ``State-space architecture of the partitioned-block-based acoustic echo controller,'' 2014, pp. 1295--1299.

\bibitem{guerin2004nonlinear}
A.~Gu{\'e}rin, G.~Faucon, and R.~Le~Bouquin-Jeann{\`e}s, ``Nonlinear acoustic echo cancellation based on volterra filters,'' \emph{IEEE Transactions on Speech and Audio Processing}, vol.~11, no.~6, pp. 672--683, 2004.

\bibitem{halimeh2019neural}
M.~M. Halimeh, C.~Huemmer, and W.~Kellermann, ``A neural network-based nonlinear acoustic echo canceller,'' \emph{IEEE Signal Processing Letters}, vol.~26, no.~12, pp. 1827--1831, 2019.

\bibitem{patel2024nonlinear}
V.~Patel, S.~S. Bhattacharjee, J.~R. Jensen, M.~G. Christensen, and J.~Benesty, ``Nonlinear acoustic echo cancellation using low-complexity low-rank recursive least-squares algorithms,'' \emph{Signal Processing}, vol. 225, p. 109623, 2024.

\bibitem{yin2024nonlinear}
K.-L. Yin, M.~M. Halimeh, Y.-F. Pu, L.~Lu, and W.~Kellermann, ``Nonlinear acoustic echo cancellation based on pipelined hermite filters,'' \emph{Signal Processing}, vol. 220, p. 109470, 2024.

\bibitem{DBLP:conf/icassp/ZhangLZ22}
C.~Zhang, J.~Liu, and X.~Zhang, ``A complex spectral mapping with inplace convolution recurrent neural networks for acoustic echo cancellation,'' in \emph{ICASSP}, 2022, pp. 751--755.

\bibitem{DBLP:journals/taslp/ZhangLL023}
C.~Zhang, J.~Liu, H.~Li, and X.~Zhang, ``Neural multi-channel and multi-microphone acoustic echo cancellation,'' \emph{{IEEE} {ACM} Trans. Audio Speech Lang. Process.}, vol.~31, pp. 2181--2192, 2023.

\bibitem{zhang2022multi}
G.~Zhang, L.~Yu, C.~Wang, and J.~Wei, ``Multi-scale temporal frequency convolutional network with axial attention for speech enhancement,'' in \emph{ICASSP}, 2022, pp. 9122--9126.

\bibitem{9413585}
N.~Howard, A.~Park, T.~Z. Shabestary, A.~Gruenstein, and R.~Prabhavalkar, ``A neural acoustic echo canceller optimized using an automatic speech recognizer and large scale synthetic data,'' in \emph{ICASSP}, 2021, pp. 7128--7132.

\bibitem{9746272}
H.~Zhao, N.~Li, R.~Han, L.~Chen, X.~Zheng, C.~Zhang, L.~Guo, and B.~Yu, ``A deep hierarchical fusion network for fullband acoustic echo cancellation,'' in \emph{ICASSP}, 2022, pp. 9112--9116.

\bibitem{Panchapagesan2022}
S.~Panchapagesan, A.~Narayanan, T.~Z. Shabestary, S.~Shao, N.~Howard, A.~Park, J.~Walker, and A.~Gruenstein, ``A conformer-based waveform-domain neural acoustic echo canceller optimized for asr accuracy.''\hskip 1em plus 0.5em minus 0.4em\relax International Speech Communication Association, 2022, pp. 2538--2542.

\bibitem{sridhar2021icassp}
K.~Sridhar, R.~Cutler, A.~Saabas, T.~Parnamaa, M.~Loide, H.~Gamper, S.~Braun, R.~Aichner, and S.~Srinivasan, ``Icassp 2021 acoustic echo cancellation challenge: Datasets, testing framework, and results,'' in \emph{ICASSP}.\hskip 1em plus 0.5em minus 0.4em\relax IEEE, 2021, pp. 151--155.

\bibitem{saka2023conversational}
A.~B. Saka, L.~O. Oyedele, L.~A. Akanbi, S.~A. Ganiyu, D.~W. Chan, and S.~A. Bello, ``Conversational artificial intelligence in the aec industry: A review of present status, challenges and opportunities,'' \emph{Advanced Engineering Informatics}, vol.~55, p. 101869, 2023.

\bibitem{revach2021kalmannet}
G.~Revach, N.~Shlezinger, R.~J. Van~Sloun, and Y.~C. Eldar, ``Kalmannet: Data-driven kalman filtering,'' in \emph{ICASSP}, 2021, pp. 3905--3909.

\bibitem{yang2023low}
D.~Yang, F.~Jiang, W.~Wu, X.~Fang, and M.~Cao, ``Low-complexity acoustic echo cancellation with neural kalman filtering,'' in \emph{ICASSP}, 2023, pp. 1--5.

\bibitem{DBLP:journals/corr/abs-2301-12363}
Y.~Zhang, M.~Yu, H.~Zhang, D.~Yu, and D.~Wang, ``Kalmannet: {A} learnable kalman filter for acoustic echo cancellation,'' \emph{CoRR}, vol. abs/2301.12363, 2023.

\bibitem{liu2023iccrn}
J.~Liu and X.~Zhang, ``Iccrn: Inplace cepstral convolutional recurrent neural network for monaural speech enhancement,'' in \emph{ICASSP}, 2023, pp. 1--5.

\bibitem{habets2006room}
E.~A. Habets, ``Room impulse response generator,'' \emph{Technische Universiteit Eindhoven, Tech. Rep}, vol.~2, no. 2.4, p.~1, 2006.

\bibitem{DBLP:conf/interspeech/SunYZH21}
Y.~Sun, L.~Yang, H.~Zhu, and J.~Hao, ``Funnel deep complex u-net for phase-aware speech enhancement,'' in \emph{Interspeech 2021}, H.~Hermansky, H.~Cernock{\'{y}}, L.~Burget, L.~Lamel, O.~Scharenborg, and P.~Motl{\'{\i}}cek, Eds.\hskip 1em plus 0.5em minus 0.4em\relax {ISCA}, 2021, pp. 161--165.

\bibitem{DBLP:journals/taslp/LuoM19}
Y.~Luo and N.~Mesgarani, ``Conv-tasnet: Surpassing ideal time-frequency magnitude masking for speech separation,'' \emph{{IEEE} {ACM} Trans. Audio Speech Lang. Process.}, vol.~27, no.~8, pp. 1256--1266, 2019.

\bibitem{10447755}
S.~He, H.~Zhang, W.~Rao, K.~Zhang, Y.~Ju, Y.~Yang, and X.~Zhang, ``Hierarchical speaker representation for target speaker extraction,'' in \emph{ICASSP}, 2024, pp. 10\,361--10\,365.

\bibitem{zhao2024attention}
F.~Zhao and X.~Zhang, ``Attention-enhanced short-time wiener solution for acoustic echo cancellation,'' \emph{arXiv preprint arXiv:2412.18851}, 2024.

\bibitem{li2021importance}
A.~Li, C.~Zheng, R.~Peng, and X.~Li, ``On the importance of power compression and phase estimation in monaural speech dereverberation,'' \emph{JASA express letters}, vol.~1, no.~1, 2021.

\bibitem{AEC-challange}
\BIBentryALTinterwordspacing
Microsoft., ``(2023) icassp acoustic echo cancellation challenge.'' [Online]. Available: \url{https://www.microsoft.com/en-us/research/academic-program/acoustic-echo-cancellation-challenge-icassp-2023/}
\BIBentrySTDinterwordspacing

\bibitem{allen1979image}
J.~B. Allen and D.~A. Berkley, ``Image method for efficiently simulating small-room acoustics,'' \emph{The Journal of the Acoustical Society of America}, vol.~65, no.~4, pp. 943--950, 1979.

\bibitem{DBLP:conf/icdsp/JeubSV09}
M.~Jeub, M.~Sch{\"{a}}fer, and P.~Vary, ``A binaural room impulse response database for the evaluation of dereverberation algorithms,'' in \emph{16th International Conference on Digital Signal Processing, {DSP} 2009}.\hskip 1em plus 0.5em minus 0.4em\relax {IEEE}, 2009, pp. 1--5.

\bibitem{DBLP:journals/corr/KingmaB14}
D.~P. Kingma and J.~Ba, ``Adam: {A} method for stochastic optimization,'' in \emph{3rd International Conference on Learning Representations, {ICLR} 2015, San Diego, CA, USA, May 7-9, 2015, Conference Track Proceedings}, Y.~Bengio and Y.~LeCun, Eds., 2015.

\bibitem{rix2001perceptual}
A.~W. Rix, J.~G. Beerends, M.~P. Hollier, and A.~P. Hekstra, ``Perceptual evaluation of speech quality (pesq)-a new method for speech quality assessment of telephone networks and codecs,'' in \emph{ICASSP}, vol.~2, 2001, pp. 749--752.

\bibitem{vincent2006performance}
E.~Vincent, R.~Gribonval, and C.~F{\'e}votte, ``Performance measurement in blind audio source separation,'' \emph{IEEE transactions on audio, speech, and language processing}, vol.~14, no.~4, pp. 1462--1469, 2006.

\end{thebibliography}

\end{document}